\documentclass[aps,prd,amsmath,amssymb,superscriptaddress,twocolumn,
nofootinbib,preprintnumbers]{revtex4-1}
\usepackage{graphicx,hyperref,color, soul,tikz}
\usepackage{cleveref,units,ctable,braket,xspace}
\usepackage{verbatim}

\hypersetup{ 
    pdfnewwindow=true,      
    colorlinks=true,       
    linkcolor=blue,         
    citecolor=blue,        
    filecolor=blue,      
    urlcolor=blue        
}  

\crefname{figure}{Fig.}{Figs.}
\crefname{equation}{Eq.}{Eqs.}
\crefname{table}{Table}{Tables}
\crefname{chapter}{Chapter}{Chapters}
\crefname{section}{Section}{Sections}
\crefname{appendix}{Appendix}{Appendices}

\usepackage{xspace}
\usepackage{slashed}

\newcommand{\lhcb}{LHCb\xspace}

\newcommand{\eg}{{\it e.g.}\xspace}
\newcommand{\ie}{{\it i.e.}\xspace}

\newcommand{\helip}{\lambda_p}
\newcommand{\helipp}{\lambda_{p^\prime}}
\newcommand{\helig}{\lambda_\gamma}
\newcommand{\helij}{\lambda_\psi}
\newcommand{\ALL}{\ensuremath{\textup{A}_{\textup{LL}}}\xspace}
\newcommand{\KLL}{\ensuremath{\textup{K}_{\textup{LL}}}\xspace}
\newcommand{\AKLL}{\ensuremath{\textup{A(K)}_{\textup{LL}}}\xspace}
\newcommand{\brpsi}{\ensuremath{\mathcal{B}_{\psi p}}\xspace}
\newcommand{\brpsiof}[1]{\ensuremath{\mathcal{B}_{\psi p}}^{#1}\xspace}

\newcommand{\GlueX}{\textsc{GlueX}}

\newcommand{\jpsi}{\ensuremath{J/\psi}\xspace}

\newcommand{\jpsip}{\ensuremath{J/\psi\,p}\xspace}

\newcommand{\nn}{\nonumber}

\newcommand{\mev}{\ensuremath{{\mathrm{\,Me\kern -0.1em V}}}\xspace}
\newcommand{\gev}{\ensuremath{{\mathrm{\,Ge\kern -0.1em V}}}\xspace}
\newcommand{\nspgev}{\ensuremath{{\mathrm{Ge\kern -0.1em V}}}\xspace}
\newcommand{\tev}{\ensuremath{{\mathrm{\,Te\kern -0.1em V}}}\xspace}

\newcommand{\Ie}{\ensuremath{I_e}\xspace}
\newcommand{\Xnot}{\ensuremath{X_0}\xspace}

\usepackage{mfirstuc} 
\newcommand{\addReviewer}[2]{
  \expandafter\newcommand\csname #1\endcsname[1]{{\bf \color{#2} \capitalisewords{#1}:\,##1}}
  \expandafter\newcommand\csname #1cor\endcsname[2]{{\color{#2} \capitalisewords{#1}:\,\st{##1}{\bf ##2}}}
  \expandafter\newcommand\csname #1color\endcsname{#2}
}

\usepackage{tikz, marginnote} 
\newcommand{\checkedby}[1]{
\ifdefined\CROSSCHECKS
  \marginnote{
    \begin{tikzpicture}
      \foreach \x [count=\xi] in {#1} {
         \node[shape=circle,inner sep=0mm,
         minimum size=2mm,
         fill=\csname \x color\endcsname] at (\xi*3mm,0) {};
       }
    \end{tikzpicture}
  }
\else
\fi
}

\usepackage{soul,color}
\definecolor{chromeyellow}{rgb}{1.0, 0.65, 0.0}
\definecolor{DodgeBlue}{rgb}{0.118, 0.565,1.000}
\definecolor{asparagus}{rgb}{0.53, 0.66, 0.42}
\definecolor{cadmiumgreen}{rgb}{0.0, 0.42, 0.24}
\definecolor{cadmiumgreen}{rgb}{0.0, 0.42, 0.24}
\definecolor{ashgrey}{rgb}{0.7, 0.75, 0.71}

\addReviewer{adam}{red}
\addReviewer{ale}{chromeyellow}
\addReviewer{vincent}{brown}
\addReviewer{misha}{cadmiumgreen}
\addReviewer{jannes}{blue}
\addReviewer{cesar}{magenta}
\addReviewer{miguel}{DodgeBlue}
\addReviewer{andrew}{purple}
\addReviewer{nathan}{orange}
\addReviewer{cris}{green}
\addReviewer{astrid}{blue}
\addReviewer{viktor}{cyan}

\usepackage{physics}

\newcommand{\jlab}{Thomas  Jefferson  National  Accelerator  Facility, Newport  News,  VA  23606,  USA}
\newcommand{\ceem}{Center for  Exploration  of  Energy  and  Matter,  Indiana  University,  Bloomington,  IN  47403,  USA}
\newcommand{\indiana}{Physics  Department,  Indiana  University,  Bloomington,  IN  47405,  USA}
\newcommand{\mainz}{Institut f\"ur Kernphysik \& PRISMA Cluster of Excellence, Johannes Gutenberg Universit\"at, D-55099 Mainz, Germany}
\newcommand{\icn}{Instituto de Ciencias Nucleares, Universidad Nacional Aut\'onoma de M\'exico, Ciudad de M\'exico 04510, Mexico}
\newcommand{\lnsmit}{Laboratory for Nuclear Science, Massachusetts Institute of Technology, Cambridge, MA 02139, USA}
\newcommand{\jleic}{Jefferson Lab, EIC Center, Newport News, VA 23606, USA}
\newcommand{\ect}{European Centre for Theoretical Studies in Nuclear Physics and Related areas (ECT$^*$)\\and Fondazione Bruno Kessler, Villazzano (Trento), I-38123, Italy}
\newcommand{\genova}{INFN Sezione di Genova, Genova, I-16146, Italy}
\newcommand{\ucm}{Departamento de F\'isica Te\'orica, Universidad Complutense de Madrid, 28040 Madrid, Spain}

\begin{document}
\preprint{JLAB-THY-19-3004}
\title{Double Polarization Observables in Pentaquark Photoproduction}
\author{D.~\surname{Winney}}
\email{dwinney@iu.edu}
\affiliation{\ceem}
\affiliation{\indiana}

\author{C.~Fanelli}
\email{cfanelli@mit.edu}
\affiliation{\lnsmit}
\affiliation{\jleic}

\author{A.~\surname{Pilloni}}
\email{pillaus@jlab.org}
\affiliation{\ect}
\affiliation{\genova}

\author{A.~N.~\surname{Hiller Blin}}
\email{hillerbl@uni-mainz.de}
\affiliation{\mainz}

\author{C.~\surname{Fern\'andez-Ram\'irez}}
\affiliation{\icn}

\author{M.~\surname{Albaladejo}}
\affiliation{\jlab}

\author{V.~Mathieu}
\affiliation{\ucm}

\author{V.~I.~\surname{Mokeev}}
\affiliation{\jlab}

\author{A.~P.~\surname{Szczepaniak}}
\affiliation{\ceem}
\affiliation{\indiana}
\affiliation{\jlab}

\collaboration{Joint Physics Analysis Center}
\noaffiliation


\begin{abstract}
We investigate the properties of the hidden-charm pentaquark-like resonances first observed by 
 LHCb in 2015, by measuring the polarization transfer \KLL between the incident photon and the outgoing proton in the exclusive photoproduction of \jpsi
near threshold. 
We present a first estimate of the sensitivity of this observable to the pentaquark photocouplings and hadronic branching ratios, and extend our predictions to the case of the initial-state helicity correlation \ALL, using a polarized target. These results serve as a benchmark for the SBS experiment at Jefferson Lab, which proposes to measure for the first time the helicity correlations \ALL\ and \KLL\ in \jpsi exclusive photoproduction, in order to determine the pentaquark photocouplings and branching ratios.
\end{abstract}
\date{\today}
\maketitle

\section{Introduction}

The LHCb data on $\Lambda_b \to J/\psi\,p\,K^-$ decay potentially indicate the existence of baryon resonances in the \jpsip spectrum~\cite{Aaij:2015tga,Aaij:2016phn,Aaij:2019vzc} that do not fit predictions of the valence quark model. 
     These would indeed
     have the minimum constituent quark content of  $uudc\bar{c}$, \ie\ that of 
     compact  hidden-charm pentaquarks or meson-baryon molecules. 
    The first partial-wave analysis of the LHCb data favored 
     two resonance
     structures, which were  labeled   $P_c(4380)$ and $P_c(4450)$. 
          The asymmetries in the angular distributions suggest that
       these two $P_c$ states~\cite{Jurik:2016bdm} have opposite parity and the 
      preferred assignments are $J^P = 3/2^-$ and $J^P= 5/2^+$ for the lighter and heavier state, respectively, but other assignments were not  ruled out. 
 
      The most recent LHCb results~\cite{Aaij:2019vzc}, however, indicate that 
       the narrower $P_c(4450)$ peak may represent two interfering states, labeled  $P_c(4440)$ and $P_c(4457)$, having widths of $<49$ and $<20 \mev$, respectively. In addition, another narrow resonance was identified at  $4312\mev$ (see also the discussion in Ref.~\cite{Fernandez-Ramirez:2019koa}). 
Given that the identification of newer   peaks came  
from the analysis of the \jpsip mass spectrum alone, the spin-parity  of these states is not  known yet. 
 For the same reason, the latter fits  do not shed more light upon the broader $P_c(4380)$ state. 

In addition to being compact five-quark states~\cite{Maiani:2015vwa,Lebed:2015tna,Anisovich:2015cia,Ali:2019npk}, alternative structures are possible.  In particular as the peaks appear close to open meson-baryon thresholds it is
 likely that they are due 
  to attractive interactions between the two hadrons.  
  For example, in the region of the  $P_c(4450)$ there 
  could be weakly bound $\bar{D}^* \Sigma_c$ and $\bar{D}^*\Sigma_c^*$ states ~\cite{Chen:2015loa,Chen:2015moa,Roca:2015dva,Guo:2019fdo,Guo:2019kdc,Xiao:2019aya,Liu:2019tjn}, and  even  in the absence of resonance or bound states it is 
   possible to generate peaks from nearby cross-channel exchanges~\cite{Szczepaniak:2015hya,Meissner:2015mza,Mikhasenko:2015vca,Guo:2016bkl}. Such ambiguities in the interpretation highlight the need for additional measurements especially with different beam-on-target configurations. 
  The use of 
  photoproduction~\cite{Kubarovsky:2015aaa,Wang:2015jsa,Karliner:2015voa,Blin:2016dlf,Wang:2019krd} is especially appealing since it reduces the role of kinematic effects  
  and minimizes model dependence of the partial-wave analysis.
  Furthermore, photoproduction at high energies is an efficient process for charm production~\cite{Chekanov:2002xi,Aktas:2005xu}, while production near threshold has long been advertised as a tool 
    for studies of the residual QCD interactions between charmonium and the nucleon~\cite{Brodsky:2000zc, Perevalova:2016dln}.
    
  The search for the $P_c(4450)$ --the narrower of the first two LHCb candidates-- 
   through a scan of the photoproduction  cross sections has been proposed by the Hall~C, CLAS12, and {\GlueX} experiments at JLab~\cite{Meziani:2016lhg,claspentaquark,Ali:2019lzf}. The first results from {\GlueX} are already available, and there is no evidence of narrow peaks~\cite{Ali:2019lzf}. Recently, an update on photoproduction studies based on the most recent LHCb results has been performed~\cite{Wang:2019krd,Cao:2019kst,Wu:2019adv},
   albeit using the spin-parity assignment of the older LHCb amplitude analysis.
   Furthermore, the use of polarization observables has been 
    recently proposed for an experiment
      at the Super BigBite Spectrometer (SBS) in Hall A at JLab~\cite{SBS:2018}. It has been argued that these may reach higher signal-to-background ratios than the usual study of differential cross sections, at least in certain parts of the parameter space, and the discovery of a double-peak structure in the $P_c(4450)$ region makes these experiments even more  relevant.
     
     In this paper we detail the 
     study of polarization observables 
      to access the pentaquark signals.
The polarization observables are sensitive to the interference between resonant and nonresonant contributions as well as between different resonance states. Polarization observables are determined  by the photoproduction amplitudes of different helicities for the initial photons, while the unpolarized cross sections are determined by the squared absolute values of the photoproduction amplitudes.
      Therefore, the polarization data offers new information that is relevant in the evaluation of the resonance photo- and hadronic couplings and it is helpful in accessing the contributions from overlapping resonances. The polarization observables extend our capabilities to validate the mechanisms of the reaction models used in the data analyses through a combined fit of unpolarized cross sections and polarization measurements.
       Here we specifically  focus on accessing 
       the sensitivity needed to investigate the properties of the pentaquarks, by studying 
        the helicity correlations between the polarized photon beam and the polarized target (\ALL) or recoil (\KLL) proton.  The latter can be assessed by measuring the polarization transfer with the one-arm polarimeter in Hall A at JLab~\cite{SBS:2018}.
        Given that there is no spin-parity assignment for 
         the new $P_c(4440)$ and $P_c(4457)$ states, which is essential for making  photoproduction predictions, and that the 
        resolution might prevent the distinction between the two, we use the previous $P_c(4450)$ information in this   feasibility study. In the following, by $P_c(4450)$ we refer to the collective effect of both $P_c(4440)$ and $P_c(4457)$ peaks. 
 We also use the information on the broad $P_c(4380)$ state, while disregarding the new $P_c(4312)$, since its spin-parity is unknown, although a similar  study can be applied in this lower mass region.  
        If photoproduction experiments prove to be successful in identifying the $P_c$ signals, an amplitude analysis  of spin-dependent observables  
         will be mandatory, for which this paper lays the groundwork. 

The paper is organized as follows. In Section~\ref{model}, we describe the reaction model for $J/\psi$ photoproduction  off the proton. In Section~\ref{results}, we
show the fits to the data and the
predictions for the 
\KLL and \ALL asymmetries
for different $P_c$ spin-parity assignments
and values of the photocouplings.  Section~\ref{sec:sensitivity} focuses on sensitivity
studies for measuring these asymmetries at Hall A of JLab. Finally,
Section~\ref{sec:summary} summarizes our conclusions.
\section{Reaction model}\label{model}
Starting from the reaction model of Ref.~\cite{Blin:2016dlf} we  
  incorporate  spin-dependent interactions
   at energies near the threshold for \jpsip 
   production. Furthermore we incorporate  both a narrow peak,  compatible with the original $P_c(4450)$ state, and the broader $P_c(4380)$.

\subsection{Background contribution}

\begin{figure}
    \centering
    \includegraphics[width=.2\textwidth]{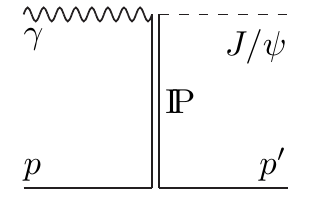}\hfill
    \includegraphics[width=.2\textwidth]{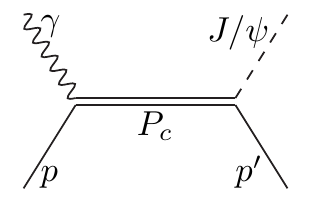}
    \caption{Relevant processes considered for near-threshold \jpsi photoproduction off proton targets. The $t$-channel process on the left describes the background, while the $s$-channel diagram to the right describes the resonant contributions from pentaquarks.}
    \label{fig:photprod}
\end{figure}
The dominant nonresonant contribution, as shown in Fig.~\ref{fig:photprod}, is assumed to be that of diffractive photoproduction of the \jpsi off the proton target. This is taken as the main 
 background to the $P_c$ signals and it is realized by an effective $t$-channel Pomeron exchange model~\cite{Close:1999bi}.   The kinematic factors and spin dependence in the model correspond to a vector exchange, to enforce that the Pomeron has an intercept which is close to unity~\cite{Lesniak:2003gf}. The resulting covariant amplitude is given by 
\begin{align}
\mel{\lambda_\psi \lambda_{p^\prime}}{T_P}{\lambda_\gamma \lambda_p} &= F(s,t) \, \bar{u}(p_f, \lambda_{p^\prime}) \gamma_\mu u(p_i, \lambda_p) \nonumber \\
&\quad\times [ \varepsilon^\mu(p_\gamma, \helig)q^\nu - \varepsilon^\nu(p_\gamma, \helig) q^\mu]\nonumber\\
&\quad\times \varepsilon^*_\nu(p_\psi, \helij).\label{background}
\end{align}
Here, $u(p_i, \lambda_p)$ and $u(p_f, \lambda_{p^\prime})$ are the Dirac spinors for the target and recoil protons, respectively, and $q$ is the photon 4-momentum. The initial and final nucleon  momenta are denoted by $p$ and $p'$ and their helicities by $\lambda_{p}$ and $\lambda_{p'}$, respectively.  The vectors $\varepsilon$  determine polarization of the photon and the \jpsi. As expected, in the high-energy limit the 
 amplitude in Eq.~(\ref{background}) is proportional to $s$, the center-of-mass energy squared (see also Appendix~C of~\cite{Mathieu:2018xyc}).
 To account for the full dependence on the Mandelstam variables $s$ and $t$ associated with  the  Pomeron trajectory, the amplitude in Eq.~({\ref{background}}) contains the function 
	 \begin{equation}\label{norm}
     F(s,t) = i A \; \bigg (\frac{s- s_{\textrm{th}}}{s_0}\bigg )^{\alpha(t)} \frac{e^{b_0(t-t_{\textrm{min}})}}{s},
     \end{equation}
where $\alpha(t) = \alpha_0 + \alpha^\prime t$ is the Pomeron trajectory (see \eg Ref.~\cite{Gribov:2009zz}). 
We fix the energy scale parameter to $s_0 = 1 \gev^2$ and $s_\textrm{th}$ to the physical threshold, $(M_\psi + M_p)^2$, where $M_\psi$ and $M_p$ are the masses of the \jpsi and the proton, respectively. 
We note that $F(s,t)$ exponentially falls when moving away from the forward direction $t = t_{\textrm{min}}$. With this parametrization, the background can be computed for the entire range of $t$ (\ie\ center-of-mass scattering angle $0^\circ \leq \theta_{\textrm{CM}} \leq 180^\circ $), but it has been derived and is more reliable in the forward region.

\subsection{Pentaquark Resonances}
Following Refs.~\cite{Blin:2016dlf,Blin:2018dkm}, 
  we parametrize 
 the pentaquark candidate contributions to  $\gamma p \rightarrow P_c \rightarrow \jpsip$ 
  using Breit-Wigner amplitudes, as shown in Fig.~\ref{fig:photprod}. This parametrization was successfully used 
   in studies of nucleon resonance photo- and electroexcitation amplitudes from the CLAS  
    data~\cite{Mokeev:2012vsa,Golovatch:2018hjk,Mokeev:2018zxt}. In terms of helicity amplitudes, 
\begin{equation} \label{resonance}
\begin{split}
\mel{\helij \helipp}{T_R}{\helig \helip} &= \\
 =f_{\textrm{th}}(s)&\frac{\mel{\helij \helipp}{T_{\textrm{dec}}}{\lambda_R} \mel{\lambda_R}{T_{\textrm{em}}^\dagger}{\helig\helip}}{M_R^2 - s - i \Gamma_RM_R}.
\end{split}
\end{equation}
We use the resonance mass $M_R$ and decay width $\Gamma_R$ for either pentaquark state as extracted from the original LHCb fit~\cite{Aaij:2015tga}. 
Since Eq.~\eqref{resonance} is finite at threshold for an $S$-wave resonance decay, but the background~ in Eq.~\eqref{norm} vanishes, we include an additional factor,
\begin{equation} \label{th-factor}
f_{\textrm{th}}(s) = \bigg ( \frac{s- s_{\textrm{th}}}{s} \frac{M_R^2}{M_R^2 - s_{\textrm{th}}} \bigg ) ^\beta,
\end{equation}
to reproduce the physical behavior of the resonant amplitude near threshold. Specifically, we choose $\beta = 3/2$, which allows the resonance signal to fall off sufficiently fast from the peak towards threshold. The numerator in Eq.~\eqref{resonance} is the product of two amplitudes. The first one,
\begin{equation}\label{hadronic}\mel{\helij \helipp}{T_{\textrm{dec}}}{\lambda_R} = g_{\helij \helipp}\!(p) \, d^{J_R}_{\lambda_R, \helij-\helipp}(\cos\theta_\text{CM}),\end{equation}
describes the coupling of the resonant state, with spin $J_R$
and helicity $\lambda_R$, to the \jpsip final state. The helicity amplitudes $g_{\helij \helipp}\!(p)$ have a near-threshold behavior $\propto p^l$ for a given decay moment $p$ and orbital angular momentum $\ell$. In general, these amplitudes depend on the final-state helicities. 
However, since nothing is known about their behavior for any of the pentaquark states, we consider them to be equal in magnitude  for any helicity: $g_{\helij \helipp}\!(p) \equiv g p^\ell$ for $\helij - \helipp > 0$, or $g_{\helij \helipp}\!(p) \equiv \eta g  p^\ell$ for $\helij - \helipp < 0$, where $\eta = \pm 1$ corresponds to the naturality of the resonance.  We assume that the amplitude of either $P_c$ state is dominated by the lowest partial wave, such that $\ell = 0$ for a $J^P_R = 3/2^-$ resonance, $\ell = 1$ for $J^P_R = 3/2^+$ or $5/2^+$, and $\ell = 2$ for $J^P_R = 5/2^-$. The magnitude of the couplings $g$ for either pentaquark state is then constrained by the partial width $\Gamma_{\psi p}$ through
\begin{align}
\Gamma_{\psi p} =& \: \brpsi\, \Gamma_R\nn\\
=& \frac{\bar p_f}{32\pi^2 M_R^2} \frac{1}{2J_R + 1}\sum_{\lambda_R\lambda_\psi\lambda_{p^\prime}}\int\mathrm{d}\Omega\left|\mel{\helij \helipp}{T_{\textrm{dec}}}{\lambda_R}\right|^2\nn\\ 
=& \frac{\bar p_f^{2\ell+1}}{8\pi M_R^2} \frac{6 g^2}{2J_R + 1},
\label{ampwidth}
\end{align} 
where \brpsi is the branching ratio of the given $P_c$ state into $\jpsi$ and the final proton with momentum $\bar{p}_f$ evaluated at the resonance peak. We note that, in general, the hadronic couplings of the different $P_c$ states are independent. 

\begin{figure*}
\includegraphics[width=.49\textwidth]{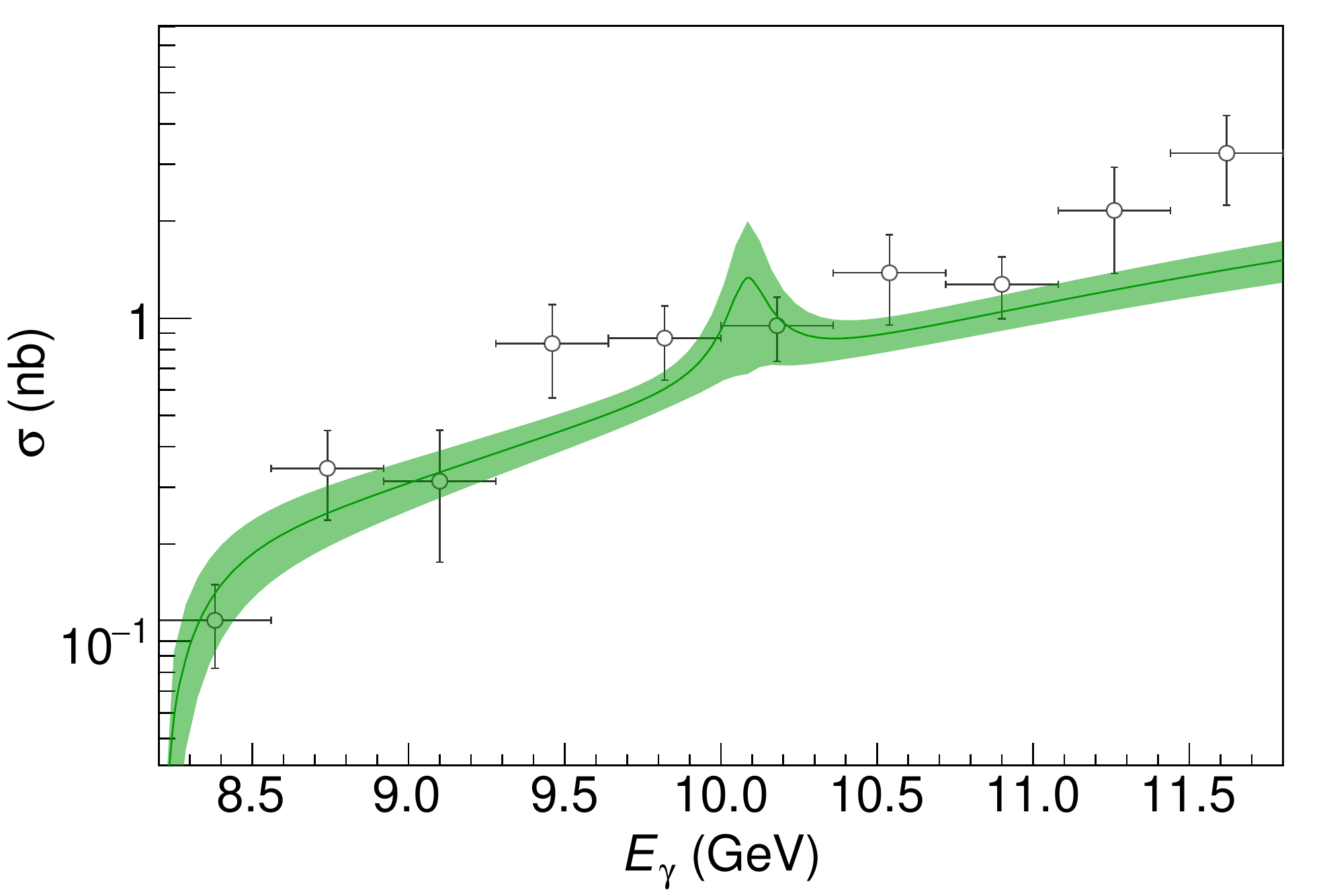}
\includegraphics[width=.49\textwidth]{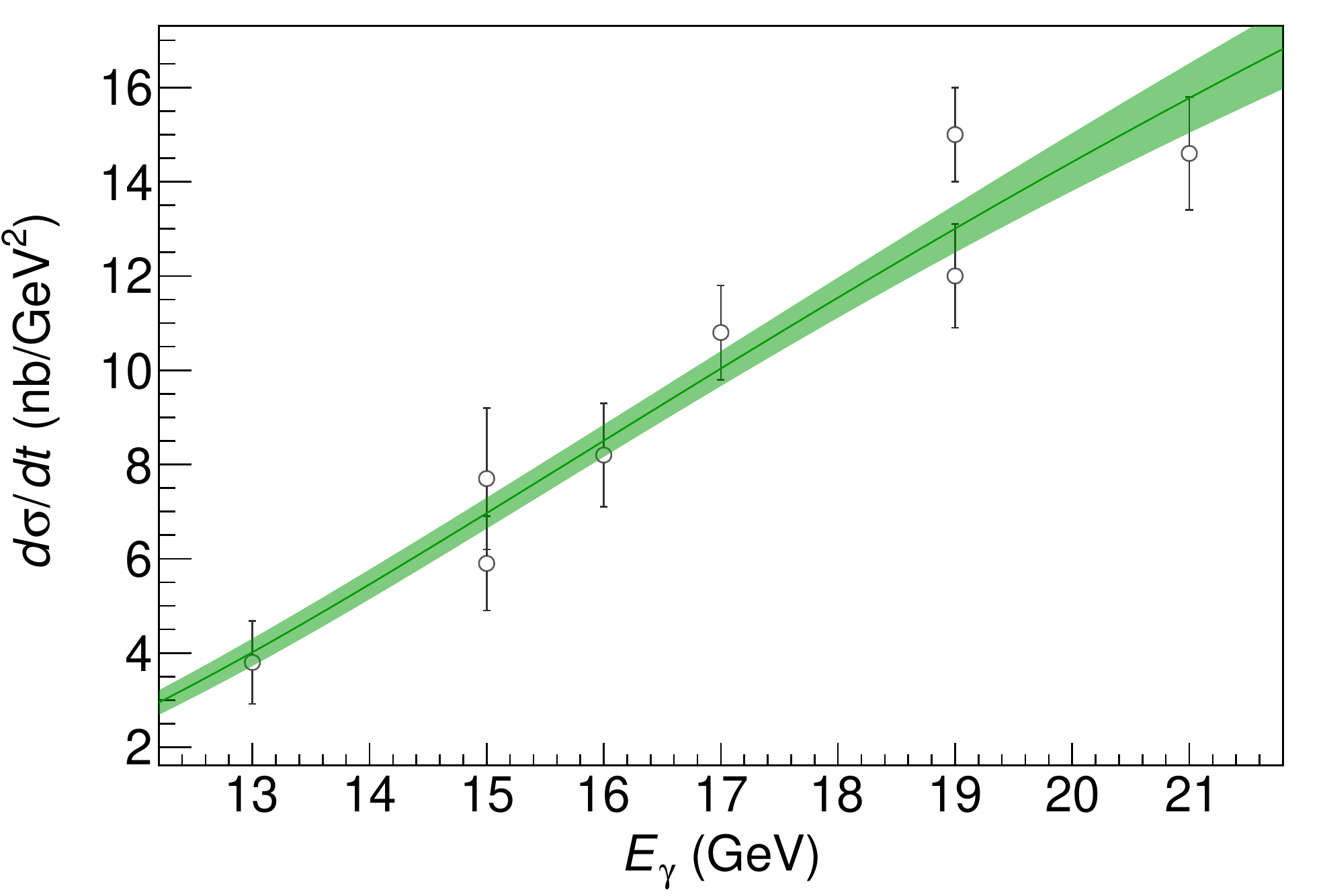}\\
\caption{Fit to \GlueX~\cite{Ali:2019lzf} (left) and SLAC~\cite{Camerini:1975cy} (right) data, for a spin assignment of the $P_c(4450)$ $J^P=\frac{3}{2}^-$. The green band represents the $1\sigma$ confidence level obtained by the bootstrap analysis.}
\label{fig:new-fit}
\end{figure*}
The second amplitude describes the photoexcitation of the pentaquark resonance, parametrized in the usual way in terms of two independent photocouplings, $A_{1/2}$ and $A_{3/2}$,
	\begin{equation} \label{em}
    \mel{\helig \helip}{T_{\textrm{em}}}{\lambda_R} = \frac{1}{M_R} \sqrt{\frac{8 s\,M_p M_R \overline{p}_i}{4\pi \alpha}} \sqrt{ \frac{\overline{p}_i}{p_i}} A_{\lambda_R}.
    \end{equation}
Here again, $\overline{p}_i$ is the momentum $p_i$ of the initial proton evaluated at the central mass of the resonance. The photocouplings are related by $A_{-\lambda_R} = \eta\, A_{\lambda_R}$.
For the electromagnetic decay width $\Gamma_\gamma$ one then obtains 
\begin{equation}
\label{a12a32width1} 
\Gamma_{\gamma}=
\frac{\bar p_i^2}
{\pi}\frac{2M_p}{(2J_{R}+1)M_{R}} \left[ \left | A_{1/2} \right |^{2}+\left |
A_{3/2} \right|^{2} \right].
\end{equation} 
As in Ref.~\cite{Blin:2016dlf}, the overall size of the photocouplings is estimated with a vector-meson dominance (VMD) model~\cite{Kubarovsky:2015aaa,Wang:2015jsa,Karliner:2015voa}, which relates the transverse \jpsi helicity amplitudes and the electromagnetic decay amplitudes through
\begin{equation}
 \bra{\lambda_{\gamma}\lambda_p}T_\text{em} \ket{\lambda_R} =
 \frac{\sqrt{4\pi \alpha} f_\psi}{M_\psi} \bra{\lambda_{\psi}=\lambda_\gamma ,\lambda_p}T_\text{dec} \ket{\lambda_R}.
 \label{vmd}
\end{equation}
 The VMD model is expected to be a robust approximation, since the quantum numbers and mass of the resonance strongly suppress resonant contributions to the electromagnetic decay other than the $\jpsi$. Using Eqs.~\eqref{ampwidth} and~\eqref{vmd}, we obtain
\begin{align}
\frac{\Gamma_\gamma}{\Gamma_{\psi p}}=4\pi \alpha  
\left(\frac{f_{\psi}}{M_{\psi}}\right)^2 \left(\frac{\bar p_i}{\bar p_f}\right)^{2\ell + 1} \mathcal{P}_t.
\label{eqvecdom}
\end{align}
The factor $\mathcal{P}_t$ is introduced to take into account that in Eq.~\eqref{vmd} only the transverse polarizations of the \jpsi contribute.  The value of $\mathcal{P}_t$ then depends on the spin-parity assignment of the pentaquark resonance, reading $\mathcal{P}_t = 2/3$ for $J^P_R = 3/2^-$, $\mathcal{P}_t = 3/5$ for $J^P_R = 5/2^+$ or $3/2^+$, and $\mathcal{P}_t = 1/3$ for $J_R^P = 5/2^-$~\cite{Kubarovsky:2015aaa}.
Combining the VMD assumption of Eq.~\eqref{eqvecdom} with Eq.~\eqref{a12a32width1}, we obtain an expression for the quadrature sum
\begin{align}
     \vert A_{1/2} \vert ^2 + \vert A_{3/2} \vert^2=&
     4\pi \alpha  
\left(\frac{f_{\psi}}{M_{\psi}}\right)^2 \left(\frac{\bar p_i}{\bar p_f}\right)^{2\ell + 1} \mathcal{P}_t\Gamma_{\psi p}\nn\\
&\times\left(\frac{\bar p_i^2}
{\pi}\frac{2M_p}{(2J_{R}+1)M_{R}}\right)^{-1},
\end{align}
which we use for the computation of the observables. Due to Eq.~\eqref{ampwidth}, it relates the photocouplings to the hadronic branching fraction size $\brpsi$.

In the fits of this work, the VMD condition is used such that $A_{1/2}=A_{3/2}$. In order to study the behavior at different relative photocoupling sizes, we then relax the equality condition, keeping the quadrature sum $|A_{1/2}|^2 + |A_{3/2}|^2$ and the size of the hadronic couplings $g$ unchanged. Thus, we define the ratio
    \begin{equation}
    \label{eq:photocoupling-ratio}
    R = \frac{A_{1/2}}{\sqrt{\vert A_{1/2} \vert ^2 + \vert A_{3/2} \vert^2}},
    \end{equation}
and treat it as a free parameter.

\section{Results}\label{results}

\begin{figure*}
\includegraphics[width=.49\textwidth]{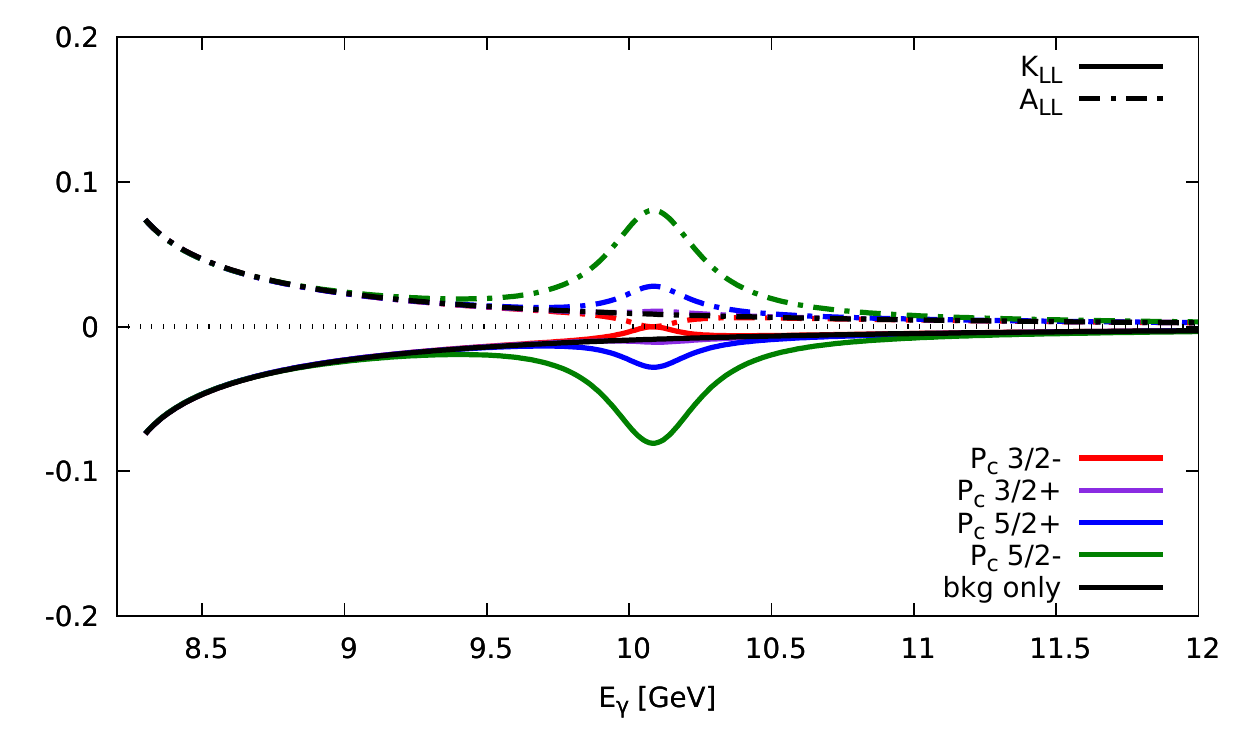}
\includegraphics[width=.49\textwidth]{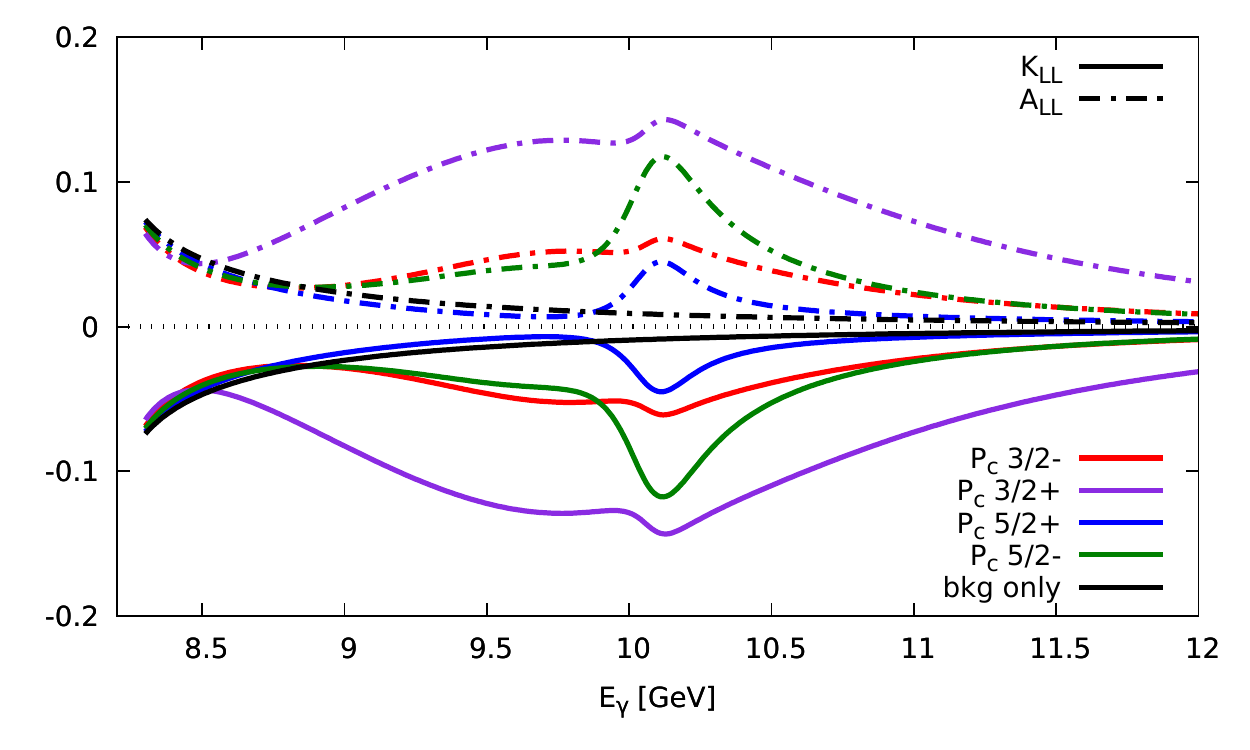}
\caption{
\ALL (dashed lines) and \KLL (solid lines) as a function of the beam energy, in the forward direction. The quoted $J^P_R$ of each colored curve is that of the $P_c(4450)$ signal. The black curves correspond to the results when no signal is included. (Left) The single pentaquark state $P_c(4450)$ is included, with equal photocouplings and $\brpsiof{(4450)} = 1\%$.
(Right) Both pentaquark states $P_c(4380)$ and $P_c(4450)$ are included. For each colored curve, the $P_c(4380)$ assumes the corresponding complementary spin-parity assignment to that of the $P_c(4450)$: the parities are opposite and the spin of the $P_c(4380)$ is 3/2 when the $P_c(4450)$ has spin 5/2 (and vice-versa). Equal photocouplings and a $\brpsi = 1\%$ for both pentaquarks are assumed.}
\label{fig:en-sp-forward-single}
\end{figure*}
\begin{figure*}
\includegraphics[width=.7\textwidth]{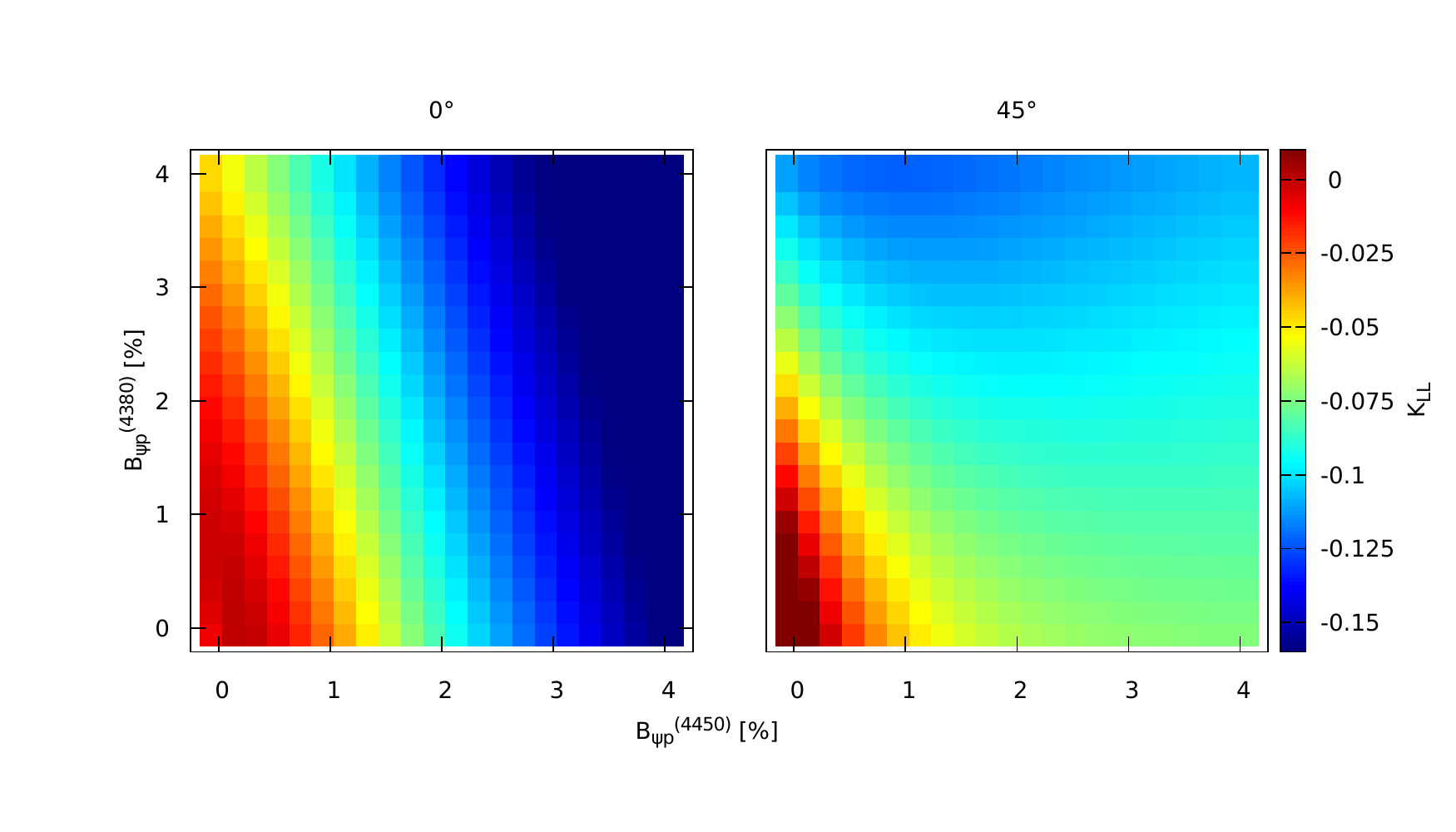}
\caption{\KLL\ as a function of the branching ratios of both pentaquark states, at the peak of the $P_c(4450)$. Here, the narrow state has $J^P = 5/2^+$ and the broader state has $J^P = 3/2^-$, and equal photocouplings for both resonances are assumed, $R^{(4450)} = R^{(4380)}=1/\sqrt{2}$. We show the results at $\theta_{\text{CM}} = 0^\circ$ (left) and $45^\circ$ (right).}
\label{fig2}
\end{figure*}
\begin{figure*}
\includegraphics[scale=.85]{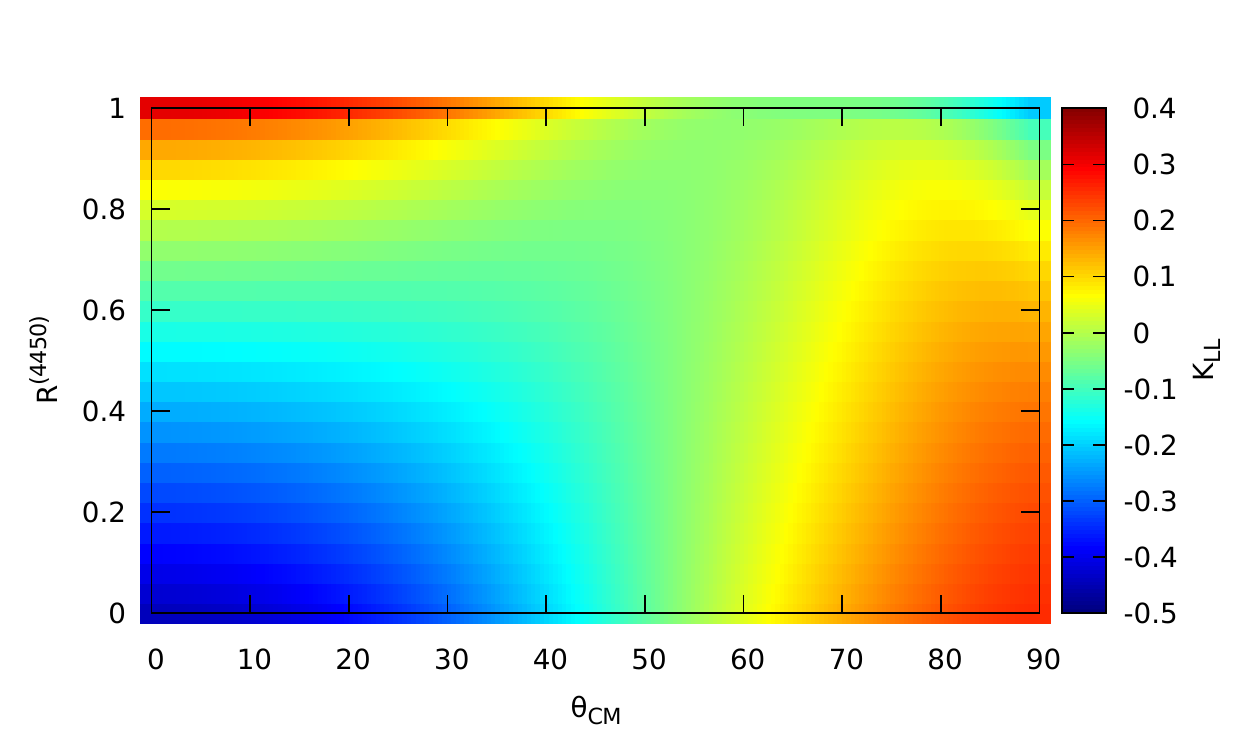}
\caption{\KLL\ dependence on the photocoupling ratio $R^{(4450)}$ and on the scattering angle $\theta_{\text{CM}}$, at the resonance energy of the narrow state.  Here, the narrow state has $J^P = 5/2^+$ and we assume $\brpsi = 1\%$ for both pentaquark states.
}\label{fig3}
\end{figure*}

While there are some quark model studies on the pentaquark photocouplings~\cite{Wang:2016dzu,Ortiz-Pacheco:2018ccl}, the experimental study of both the pentaquark hadronic and photocouplings is \textit{terra incognita}.
The spin-dependent observables \ALL (\KLL) describe correlations between the helicities of the incoming photon and the incoming (outgoing) proton, given in terms of differential cross sections as
\begin{equation}\label{eq:all}
\AKLL=   \frac{1}{2} \left[\frac{d\sigma(++)-d\sigma(+-)}{d\sigma(++)+d\sigma(+-)} - \frac{d\sigma(-+)-d\sigma(--)}{d\sigma(-+)+d\sigma(--)}\right],
\end{equation}
with $d\sigma\equiv d\sigma/d \cos\theta_\text{CM}$, where the first helicity refers to the 
incident photon, and the second refers to the target (\ALL) or recoil (\KLL) proton in the center-of-mass frame~\cite{Kroll:2018uvl}. 
For the computation of \ALL, the differential cross sections are obtained from the helicity amplitudes as follows
\begin{align}
    d\sigma(\lambda_\gamma \lambda_p)&= \frac{ 4\pi\alpha}{32 \pi  s} \frac{p_f}{p_i}  
\sum_{\lambda_{\psi}, \lambda_{p'}}
\left|\bra{\lambda_{\psi}\lambda_{p^\prime}} T \ket{\lambda_\gamma \lambda_p}\right|^2\text{ for \ALL}\nn,\\
    d\sigma(\lambda_\gamma \lambda_{p'})&= \frac{ 4\pi\alpha}{32 \pi  s} \frac{p_f}{p_i}  \frac{1}{2}
\sum_{\lambda_{\psi}, \lambda_{p}}
\left|\bra{\lambda_{\psi}\lambda_{p'}} T \ket{\lambda_\gamma \lambda_p}\right|^2\text{ for \KLL}.
\end{align}

\begin{table*}[t]
		\caption{Parameters of the fits for different $J^P$ assignments for the $P_c(4450)$ state. Uncertainties are at the 68\% confidence level,  except for the branching ratio, whose upper limit is quoted at 95\%.}
	\begin{ruledtabular}
	\begin{tabular}{|c| c c c c |}\label{mean-fit}
	    $J^P$ 				& $\frac{3}{2}^-$ 		& $\frac{5}{2}^+$ 	& $\frac{3}{2}^+$ 		& $\frac{5}{2}^-$ \\ 
        \hline
        $A$					& $0.379\pm 0.051$		& $0.380 \pm0.053$ 	& $0.378 \pm 0.049$		& $0.381 \pm 0.053$ 		\\
        $\alpha_0$ 				& $0.941\pm 0.047$ 		& $0.941 \pm0.049$ 	& $0.942 \pm 0.045$		& $0.941 \pm 0.048$ 		\\
        $\alpha^\prime$ ($\nspgev^{-2}$)		& $0.364\pm0.037$ 		& $0.367 \pm0.039$ 	& $0.363 \pm 0.035$		& $0.365 \pm 0.037$ 		\\
        $b_0$ ($\nspgev^{-2}$)			& $0.12\pm 0.14$ 		& $0.13 \pm0.15$   	& $0.12 \pm 0.14$		& $0.13 \pm 0.15$  		 \\
        $\brpsiof{(4450)}$ (95\%) 				& $\leq 4.3\%$			& $ \leq 1.4\%$ 	& $\leq 1.8\%$ 			& $\leq 0.71\%$ 
	\end{tabular}
    \end{ruledtabular}
\end{table*}

The unpolarized differential and total cross sections are given by
\begin{subequations}
\begin{align}
\label{Edsigdcos}
\frac{d\sigma}{dt} &= \frac{ 4\pi\alpha}{64 \pi  s\,p_i^2} \frac{1}{4} \!\!
\sum_{\lambda_\gamma,\lambda_p, \lambda_{\psi}, \lambda_{p'}} \!\!
\left|\bra{\lambda_{\psi}\lambda_{p^\prime}} T \ket{\lambda_\gamma \lambda_p}\right|^2,\\
\sigma &= \int_{t_\text{max}}^{t_\text{min}} d t \frac{d\sigma}{dt},\label{sigmaintegrated}
\end{align}
\end{subequations}
where as customary $t_\text{min(max)} = M_\psi^2 - 2 p_i \left(\sqrt{p_f^2 + M_\psi^2}  \mp p_f\right)$. In Eq.~\eqref{Edsigdcos}, $T = T_P + T_R(4380) + T_R(4450)$ is the coherent sum of the Pomeron background in Eq.~\eqref{background} and the resonant contribution from the pentaquarks given by Eq.~\eqref{resonance}. Since the differential cross section is not very sensitive to a broad $P_c(4380)$, we set $\brpsiof{(4380)} = 0$ in the fits to reduce the parameter space. The resonant parameters of the $P_c(4450)$ are fixed to the LHCb best values~\cite{Aaij:2015tga}, while $\brpsiof{(4450)}$ is free. The background parameters $\alpha_0$, $\alpha^\prime$, $b_0$, and $A$ in Eq.~\eqref{background} are also fitted.
We fit \jpsi photoproduction data from \GlueX~\cite{Ali:2019lzf} and SLAC~\cite{Camerini:1975cy}. Unlike our previous works~\cite{Blin:2016dlf,Blin:2018dkm}, we ignore the very high-energy data from HERA and ZEUS~\cite{Chekanov:2002xi,Aktas:2005xu}, as well as the old unpublished data close to threshold~\cite{Ritson:1976rj,Anderson:1976sd}. This is done in order to have a better description of the region of interest which is now better constrained thanks to the \GlueX\ data.  For the background model to best reproduce the data, we include in the fit both the energy- and $t$-dependent information from Tables~I and~II of~\cite{Ali:2019lzf}. 
However, since the points come from the same data set and correlations are not reported, our statistical estimates must be considered with care. The curves are integrated over the (large) bin size, while the energy resolution is neglected.
The mean fit parameters and their uncertainties for each spin-parity assignment of the $P_c(4450)$ have been calculated employing the bootstrap technique (see~\cite{Blin:2016dlf} for details), and the results  are shown in~\cref{mean-fit,fig:new-fit}. The results for $\alpha_0$ are compatible with unity. The values of $\alpha^\prime$ are a bit higher, but marginally compatible with the ones extrapolated from the SPS energies~\cite{Erhan:1999gs}.  We use Eqs.~\eqref{background},~\eqref{resonance}, and~\eqref{eq:all} to give the predicted values of \ALL\ and \KLL\ for a given beam energy $E_\gamma$ and center-of-mass scattering angle $\theta_\text{CM}$. Note that the beam energy corresponding to the $P_c(4450)$ peak is $E_\gamma\approx10.6\gev$, while for the  $P_c(4380)$ it is $E_\gamma\approx9.8\gev$. Some predictions for the polarization observables are shown in \cref{fig:en-sp-forward-single,fig2,fig3}.

\section{Sensitivity studies}\label{sec:sensitivity}
\begin{figure}[t]
\includegraphics[width=.49\textwidth]{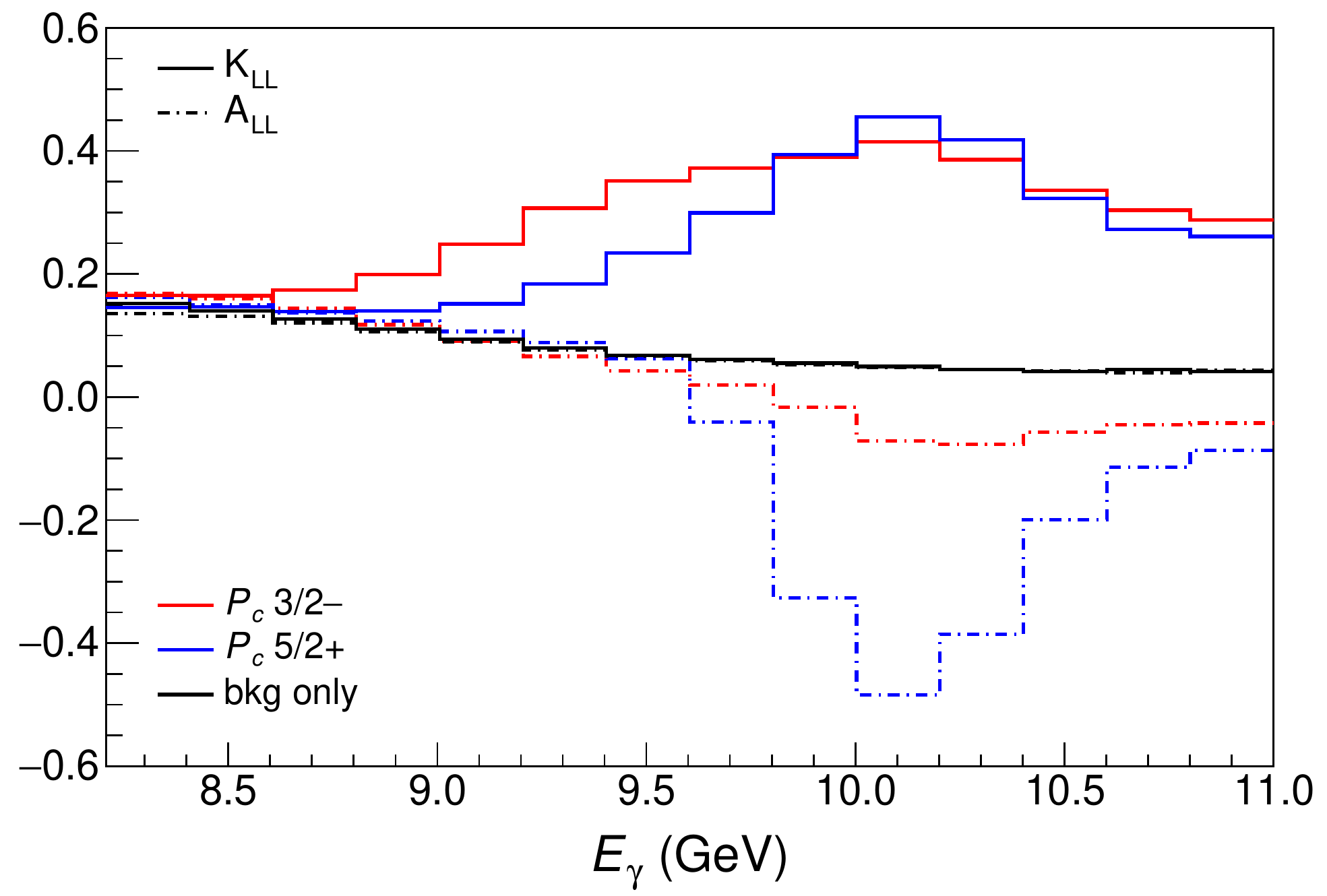}
	\caption{Predictions for \KLL (solid line) and \ALL (dash-dotted line) in the SBS acceptance in bins of energy, considering $\brpsiof{(4450)} = \brpsiof{(4380)} = 1.3\%$, $R^{(4450)} = 0.2$, $R^{(4380)} = 1/\sqrt{2}$ and a resolution of $125\mev$. The scenarios shown are for a narrow pentaquark with $J^P_R = 3/2^-$ (red), $J^P_R = 5/2^+$ (blue) and the case without pentaquarks (black). Note the sign flip of the two observables with respect to the forward prediction of \cref{fig:en-sp-forward-single}.
}
\label{fig:withacc}
\end{figure}

A measurement of the polarization observables was recently proposed 
for Hall A at JLab~\cite{SBS:2018}. 
 It will take advantage of the SBS setup developed for the GEp/SBS experiment in Hall~A~\cite{Gnanvo:2014hpa}, which consists of a hadron (SBS) and an electron arm (ECAL). To measure the initial helicity state correlation \ALL,  the experiment will scatter a circularly polarized photon beam onto a longitudinally polarized (NH$_{3}$) target. The polarization transfer \KLL\ will be measured by a proton polarimeter of the GEp setup in the SBS arm, used in combination with an unpolarized liquid hydrogen target. With this experimental setup, all three final-state particles $\gamma p \to \jpsi(\to e^+e^-) p$
 will be reconstructed, allowing for a clean signature of the candidate events with suppressed backgrounds~\cite{SBS:2018,Ali:2019lzf}. An example of the predictions for the polarization observables in the SBS acceptance is given in~\cref{fig:withacc}.

We provide a sensitivity study of the main parameters of the LHCb pentaquark states based on toy Monte Carlo simulations of \ALL and \KLL experiments at JLab. As we said, this study relies on the two states seen in 2015, and the actual sensitivity is susceptible to change when more information about the new $P_c(4440)$ and $P_c(4457)$ becomes available. The code we used to calculate the observables is publicly available on the JPAC website~\cite{JPACweb}, and can be extended to other spin-parities and resonance parameters. 
The statistical uncertainty on \ALL  and \KLL\ can be approximated according to~\cite{SBS:2018}\footnote{In the low-statistics limit, we imposed that these uncertainties cannot exceed 1.}
\begin{align}\label{uncertainty1}
\Delta \ALL &\sim \frac{1}{\sqrt{ \frac{N_{\psi}}{R}} \cdot P_{p} \cdot P_{\gamma}},
\intertext{and} 
\label{uncertainty2}
\Delta \KLL &\sim \frac{1}{\sqrt{\frac{N_{\psi}}{R}} \cdot \left\langle F_\text{eff} \right\rangle \cdot P_{\gamma}},
\end{align}
where $N_{\psi}$ is the total number of exclusive \jpsi events expected to be detected. 
The factor $R$ is the rescaling due to the (small) background underneath the \jpsi
peak,\footnote{The background comes mainly from Bethe-Heitler continuum $e^+e^-$ production.} and can be safely assumed $\approx 1$.
It is worth recalling that the pentaquark signals and the Pomeron background are summed at the amplitude level, and for this reason in the equations above we do not define the number of pentaquark events, but rather use the number 
 of measured $\jpsi \,p$ candidates. The photon beam polarization is approximately
$P_{\gamma} \sim 0.8$,\footnote{This comes from an initial electron beam polarization $P_{e} \sim 0.85$ and the Maximon-Olsen formula~\cite{Olsen:1959zz} as a function of the incident $E_{\gamma}$, ranging from the \jpsi threshold to the end-point energy $E_{e}$.} while $P_{p}\sim 0.75$ is the average target proton polarization.
In the case of \KLL, one considers the polarization transferred to the recoil proton. The average effective figure of merit $\left\langle F_\text{eff} \right\rangle=\sqrt{\epsilon_\text{pol}} A_{y} \sin \chi_\text{prec}$ includes the polarimeter efficiency $\epsilon_\text{pol}$, the polarimeter analyzing power $A_{y}$, and the spin precession angle in the SBS magnet $\chi_\text{prec}$. Following Ref.~\cite{SBS:2018}, this figure of merit is approximated as $\left\langle F_\text{eff} \right\rangle \sim 0.07$. 

\begin{table}[t]
\caption{\label{tab:exper_settings}
Values of the experimentally projected beam current \Ie, length of the radiator in terms of radiation length \Xnot and thickness times density of the target  $\rho_\text{free}\cdot l$.
}
	\begin{ruledtabular}

\begin{tabular}{ |  c| c | c | c |}
 & \Ie [$\mu\text{A}$] & \Xnot & $\rho_\text{free} \cdot l$ [$\text{g}/\text{cm}^{2}$] \\
\hline
\KLL (SBS) & 5.0 & 6$\cdot$10$^{-2}$  &   1.08  \\
\ALL (SBS) & 0.1 & 10$\cdot$10$^{-2}$ & 0.32  \\
\end{tabular}
\end{ruledtabular}

\end{table}

Following the experimental design concept of Ref.~\cite{SBS:2018}, SBS and ECAL are located at the right and left of the beamline, respectively, with central polar angles of 17$^{\circ}$ and 22$^{\circ}$. 
We consider the experimental signatures that provide the best energy and mass resolutions, $\sigma(E_{\gamma})\sim 125\mev$ and $\sigma(M_{\jpsi})\sim 20 \mev$
at an electron beam energy of $10\gev$, which is where one of the two leptons is reconstructed along with the proton in the hadronic arm, and the other lepton is detected in the electromagnetic calorimeter. 
Our minimal requirements are for the proton and the lepton in the hadronic arm to have an energy of $2$ and $1\gev$, respectively,
and the other lepton to deposit an energy of $1\gev$ in the calorimeter. 
We refer to Ref.~\cite{SBS:2018} for further details on the experimental settings and selection criteria which have also been used for the simulation studies of the present paper. The final acceptance with these cuts is $\sim 1\%$. 
The expected yields are calculated based on the experimental conditions of Table~\ref{tab:exper_settings} and by requiring the events to be within the detector acceptance. In particular, for a given photon energy range $(E_{1},E_{2})$ and time interval $\Delta t$, the yield is estimated as $N_{\psi} \approx  I_{e} \cdot \left( \int_{E_{1}}^{E_{2}} 
f(E_{\gamma}) \,\sigma(E_{\gamma}) \,dE_\gamma \right) \cdot (\rho \cdot l) \cdot  \epsilon \cdot \mathcal{B}(\psi\to e^+e^-) \cdot \Delta t$, 
where $I_{e}$ is the electron beam current, $\sigma$ is the photoproduction cross section in Eq.~\eqref{sigmaintegrated} as a function of the incident photon energy, 
$f(E_{\gamma})$ is the bremsstrahlung photon flux calculated for a radiator with \Xnot radiation lengths according to Ref.~\cite{Mo:1968cg}, $\rho\cdot l$ is the product of the target density and length, $\epsilon$ is the detection acceptance, and $\mathcal{B}\!\left(\psi\to e^{+}e^{-}\right)=5.94\%$~\cite{pdg}.
The values of \Ie, \Xnot and $\rho\cdot l$ are given in Tab.~\ref{tab:exper_settings}. These values propagate into the statistical uncertainties defined by Eqs.~\eqref{uncertainty1} and~\eqref{uncertainty2}. A fictitious systematic uncertainty of 2\% was taken into account in the toy model. The spectrum of the bremsstrahlung photons is calculated as in Ref.~\cite{Mo:1968cg}. 
The incident photon energy of interest ranges from about the \jpsi production threshold to the end-point energy coinciding with the electron beam.

The proxy used to estimate the sensitivity to the $P_{c}$ states is based on the log-likelihood difference $\Delta\log{\mathcal{L}}$ between the background-only hypothesis, and the hypothesis that two $P_{c}$ resonances interfere with it.
Wilks' theorem then relates the value of  $-2\Delta\log{\mathcal{L}}$ to a $\chi^{2}$ distribution with degrees of freedom equal to the difference in dimensionality between the two hypotheses~\cite{Wilks:1938dza}.

\begin{figure*}
\includegraphics[width=.48\textwidth]{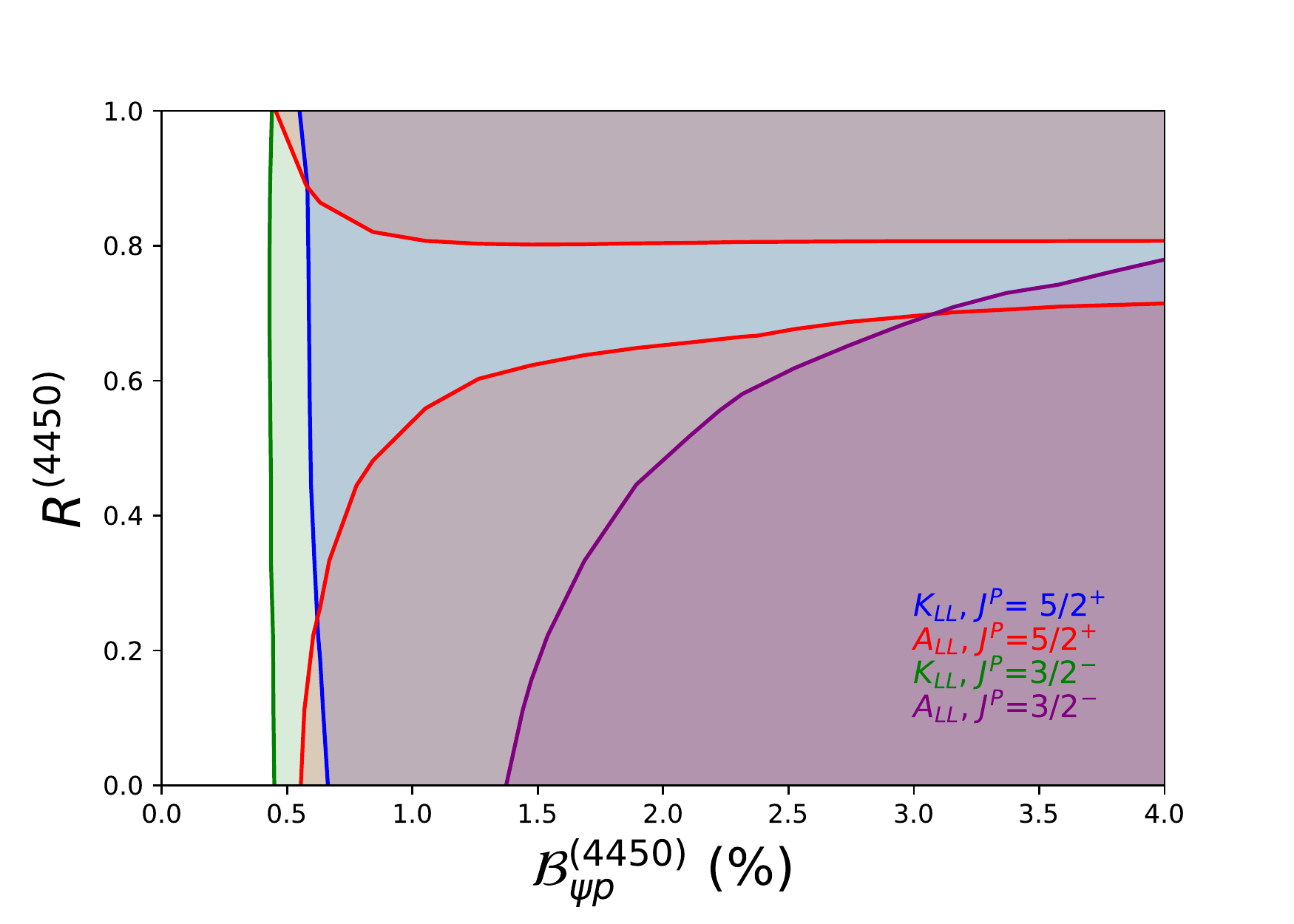}
\includegraphics[width=.48\textwidth]{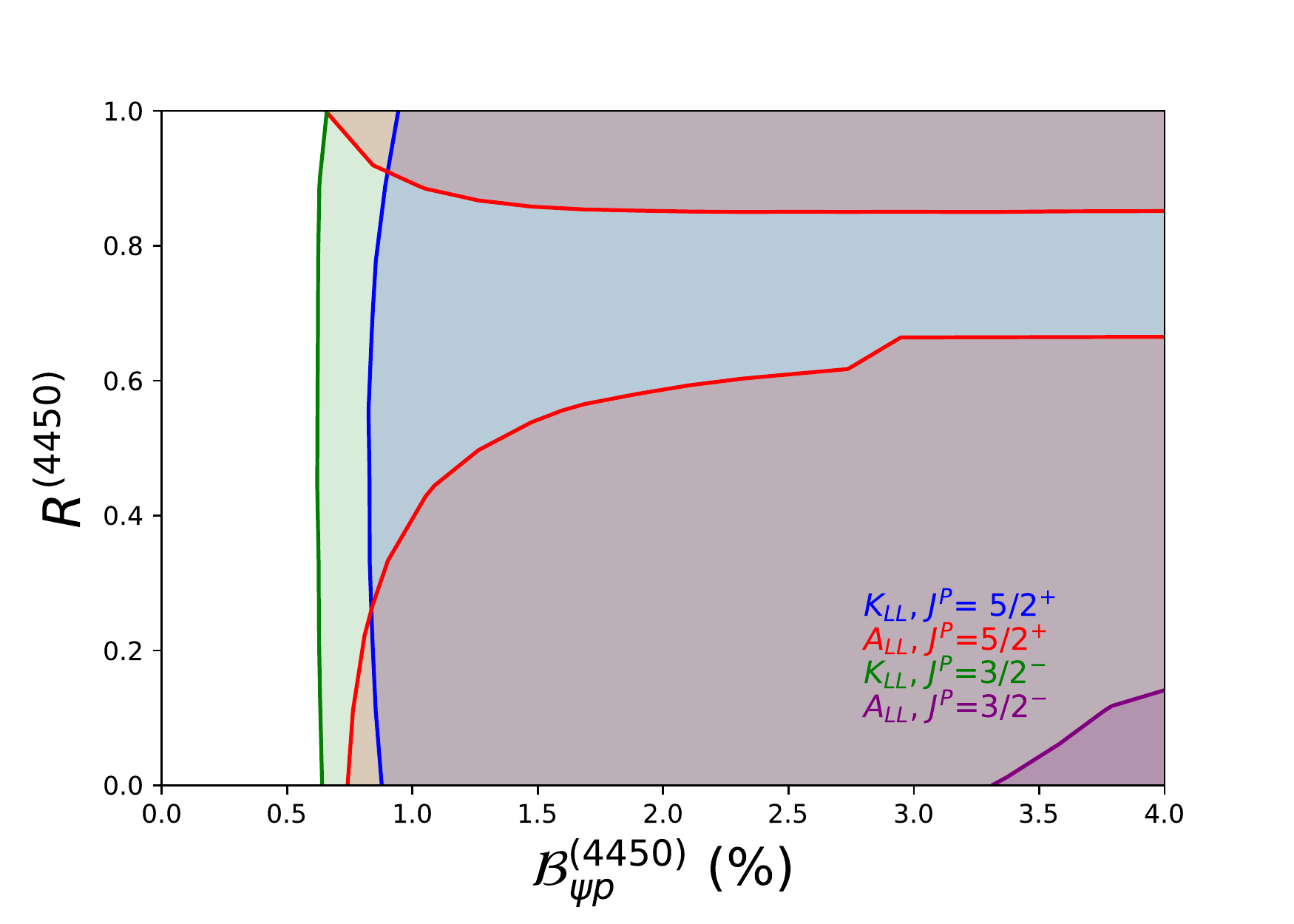}
 \caption{Sensitivity to the $P_c(4450)$ 
with spin-parity $3/2^{-}$ and $5/2^{+}$, as a function of $\brpsiof{(4450)}$, obtained from a log-likelihood analysis. For SBS we assume 250 days of data taken with the experimental settings of Table~\ref{tab:exper_settings}.  
 The colored areas highlight where the signals would be observed beyond $2\sigma$ (left plot) or $5\sigma$ (right plot).
 In the depicted scenario, the $P_c(4380)$ is assumed to have a spin-parity assignment complementary to the $P_c(4450)$ as explained in the text, equal photocouplings $R^{(4380)} = 1/\sqrt{2}$, and the same branching ratio as the $P_c(4450)$.
 }
\label{fig:sensitivity_results}
\end{figure*}

We focus here on a particular scenario, where the broad $P_c(4380)$ has a photocoupling ratio $R^{(4380)} = 1/\sqrt{2}$ (corresponding to equal photocouplings $A_{1/2}^{(4380)} = A_{3/2}^{(4380)}$), the hadronic branching ratio is equal to that of the $P_{c}(4450)$, and the mass and the width of the two states are fixed
to the best values measured in Ref.~\cite{Aaij:2015tga}.

Pseudodata are generated in a 2D grid of points, varying the photocoupling ratio and the hadronic branching ratio of the $P_c(4450)$. 
For each point of the grid, multiple $\mathcal{O}(10^{3})$ toy models are computed producing binned data of \ALL and \KLL as a function of the incident energy.

The results of the sensitivity studies can be found in Fig.~\ref{fig:sensitivity_results} for the exemplary spin-parity assignments $5/2^{+}$ and $3/2^{-}$ of the $P_c(4450)$, assuming that the $P_c(4380)$ havs opposite signature as explained above. They have been estimated assuming 250 days of collected data, both for \KLL and \ALL, and a 80\% live time. The effective efficiency includes the geometrical acceptance and a conservative detection efficiency $\epsilon_\text{reco}\sim 50\%$ to reconstruct the channel.
 We find that there is a projected sensitivity of more than  $5\sigma$ in a larger region than the one already excluded by the cross section measurements only, in particular for \KLL. 


\section{Summary}\label{sec:summary}

We presented for the first time a study of the polarization observables in hidden charm pentaquark photoproduction close to  threshold. This was motivated by a recent Letter of Intent for the SBS experiment at Hall A of JLab, which proposed to study the polarization observables \KLL\ and \ALL\, due to their higher sensitivity to the signal when compared to data on differential cross sections. 
       
       We thus analyzed the possibility of observing these exotic structures, treating the $P_c(4440)$ and $P_c(4457)$ states as one combined $P_c(4450)$ peak, since there is as of now no information on the quantum numbers of the individual states. We updated the model in~\cite{Blin:2016dlf}, considering a Pomeron-like background added coherently to the two resonances $P_c(4450)$ and $P_c(4380)$, and fit to the available data on \jpsi photoproduction close to threshold~\cite{Camerini:1975cy,Ali:2019lzf}, including the new \GlueX\ results. 
       
       If photoproduction experiments prove to be successful in pinning down the $P_c$ signals, more refined and systematic analyses on the differential cross section and the spin-parity properties of the pentaquarks will be mandatory, for which this work serves as a benchmark.
 We show that 250 days of collected data with the SBS experiment will give more than $5\sigma$ sensitivity to the $P_c$ signals in large regions of the parameter space, in particular for \KLL. 

In conclusion, the polarization observables showed an excellent sensitivity to both photo- and hadronic couplings. Therefore, they provide a way to study the nature and properties of the exotic resonances. 

The code to calculate the observables and generate the Monte Carlo toy data is publicly available on the JPAC website~\cite{JPACweb}.

\begin{acknowledgments}
We thank A.~Deur and M.~Williams for useful discussions.  We also thank L.~Pentchev and B.~Wojtsekhowski for useful comments on the manuscript.
This work was supported by
the U.S.~Department of Energy under Grants
No.~DE-AC05-06OR23177, 
No.~DE-FG02-87ER40365, 
and No.~DE-FG02-94ER40818, 
PAPIIT-DGAPA (UNAM, Mexico) Grant No.~IA101819, 
CONACYT (Mexico) Grants No.~251817
and~No.~A1-S-21389. 
This work was also supported by the Deutsche Forschungsgemeinschaft (DFG, German Research
Foundation), in part through the Collaborative Research Center [The Low-Energy Frontier of the Standard Model,
Projektnummer 204404729 - SFB 1044], and in part through the Cluster of Excellence [Precision Physics, Fundamental
Interactions, and Structure of Matter] (PRISMA+ EXC 2118/1) within the German Excellence Strategy (Project ID
39083149). 
V.M. acknowledges support from Comunidad Aut\'onoma de Madrid through 
Programa de Atracci\'on de Talento Investigador 2018 (Modalidad 1).
\end{acknowledgments}

\bibliographystyle{apsrev4-1-jpac}
\bibliography{quattro}
\end{document}